\documentclass[acmsmall, screen]{acmart}
\AtBeginDocument{%
  \providecommand\BibTeX{{%
    \normalfont B\kern-0.5em{\scshape i\kern-0.25em b}\kern-0.8em\TeX}}}

\setcopyright{acmlicensed}
\acmJournal{TOSEM}
\acmYear{2024} \acmVolume{1} \acmNumber{1} \acmArticle{1} \acmMonth{1}\acmDOI{10.1145/3664811}

\acmJournal{TOSEM}
\acmVolume{37}
\acmNumber{4}
\acmArticle{111}
\acmMonth{8}

\usepackage{amsmath,amssymb,amsfonts}
\usepackage{algorithmic}
\usepackage{graphicx}
\usepackage{textcomp}
\usepackage{xcolor}
\usepackage{balance} 
\usepackage{url} 
\def\BibTeX{{\rm B\kern-.05em{\sc i\kern-.025em b}\kern-.08em
    T\kern-.1667em\lower.7ex\hbox{E}\kern-.125emX}}
\usepackage{textcomp}
\usepackage{booktabs}
\usepackage{enumitem}
\usepackage{multirow}
\usepackage{subfig}
\usepackage{listings}
\usepackage{color}
\usepackage{threeparttable}
\usepackage{array}
\usepackage{diagbox}
\usepackage{graphicx}
\usepackage{verbatim}
\usepackage{url}
\usepackage{soul}
\soulregister\cite7
\soulregister\ref7
\usepackage{graphicx}
\usepackage{multirow}
\usepackage{hhline}
\usepackage{comment}
\usepackage{float}
\usepackage{array,booktabs}
\usepackage{marvosym} 
\usepackage{pifont}
\usepackage{tcolorbox}
\usepackage{bm}
\usepackage[ruled,linesnumbered]{algorithm2e}
\usepackage{amsthm}

\newcommand{\todo}[1]{}

\renewcommand{\todo}[1]{{\color{red} TODO: {#1}}}

\newcommand{\changed}[1]{\textcolor{black}{#1}}

\begin{document}

\title[What Makes a Good TODO Comment?]{What Makes a Good TODO Comment?}

\author{Haoye Wang}
\affiliation{%
  \institution{Hangzhou City University}
  \country{China}
  }
\email{wanghaoye@hzcu.edu.cn}

\author{Zhipeng Gao}
\authornote{This is the corresponding author}
\affiliation{%
  \institution{Shanghai Institute for Advanced Study of Zhejiang University}
  \country{China}
  }
\email{zhipeng.gao@zju.edu.cn}

\author{Tingting Bi}
\affiliation{%
  \institution{The University of Western Australia}
    \country{Australia}
  }
\email{Tingting.Bi@uwa.edu.au}

\author{John Grundy}
\affiliation{%
  \institution{Monash University}
  \city{Melbourne,}
  \state{VIC}
  \postcode{3168}
  \country{Australia}
  }
\email{john.grundy@monash.edu}

\author{Xinyu Wang}
\affiliation{%
  \institution{Zhejiang University}
  \country{China}
  }
\email{wangxinyu@zju.edu.cn}

\author{Minghui Wu}
\affiliation{%
  \institution{Hangzhou City University}
  \country{China}
  }
\email{mhwu@hzcu.edu.cn}

\author{Xiaohu Yang}
\affiliation{%
  \institution{Zhejiang University}
  \country{China}
  }
\email{yangxh@zju.edu.cn}

\renewcommand{\shortauthors}{Wang et al.}

\newcommand{\find}[1]{
\begin{tcolorbox}[leftrule=1mm,toprule=0mm,bottomrule=0mm,left=1pt,right=2pt,top=2pt,bottom=2pt] #1
\end{tcolorbox}
}

\begin{abstract}
Software development is a collaborative process that involves various interactions among individuals and teams.
TODO comments \changed{in source code} play a critical role in managing and coordinating diverse tasks during this process.
However, \changed{this study} finds that a large proportion of open-source project TODO comments are left unresolved \changed{or} take a long time to be resolved.
About 46.7\% of TODO comments in open-source repositories are of low-quality \changed{(e.g., TODOs that are ambiguous, lack information, or are useless to developers)}.
This highlights the need for better TODO practices.
In this study, we investigate four aspects regarding the quality of TODO comments in open-source projects: 
(1) the prevalence of low-quality TODO comments; 
(2) the key characteristics of high-quality TODO comments; 
(3) how are TODO comments of different quality managed in practice; 
and (4) the feasibility of automatically assessing TODO comment quality.
Examining 2,863 TODO comments from Top100 GitHub Java repositories, we propose criteria to identify high-quality TODO comments and provide insights into their optimal composition.
We discuss the lifecycle of TODO comments with varying quality. 
To assist developers, 
we construct deep learning-based methods that show promising performance in identifying the quality of TODO comments, potentially enhancing development efficiency and code quality.

\end{abstract}




\maketitle
\section{Introduction}\label{sec:intro}

Software development involves not only the interaction between activities and software artifacts, but also the interaction between developers.
In addition to source code, natural language annotations play a vital role in these two aspects~\cite{de2005study}.
\changed{\textbf{\textit{TODO comments}} are extensively used by software developers not only to manage their personal pending tasks but also to coordinate with other developers~\cite{sridhara2016automatically, nie2018natural}, thereby acting as a key mechanism to synchronize work among team members.}
For example, one of the most common usages of TODO comments is being dropped as a reminder to add or delete features~\cite{esfandiari2023exploratory}.
Some developers also use TODO comments as a request to check for a potential problem.
Developers might use TODO comments to specify testing requirements for testers, or a code reviewer might leave TODO comments indicating pending tasks that the code submitter needs to address.
We refer to a TODO comment as `TODO' and multiple TODO comments as `TODOs' in the rest of this paper. 

TODOs are often introduced with code changes, which carry valuable information about code changes that can improve the software quality, performance, maintenance, and reliability~\cite{nie2018natural}.
Ying et al.~\cite{ying2005source} identified the high frequency and widespread use of TODOs, and highlighted their significance in communicating among developers. 
In modern software development, it is common for teams to be distributed and composed of individuals from diverse cultural and educational backgrounds. 
\changed{For example, when the TODO keyword gets flagged, every developer involved in the software development process, regardless of their cultural or educational differences, knows which controls need to be reactivated before the next round of updates. By embedding these clear, actionable items directly in the code, TODOs provide a simple yet effective way to synchronize work across the development team.}








\begin{figure}[h]
\centering
\includegraphics[scale=0.47]{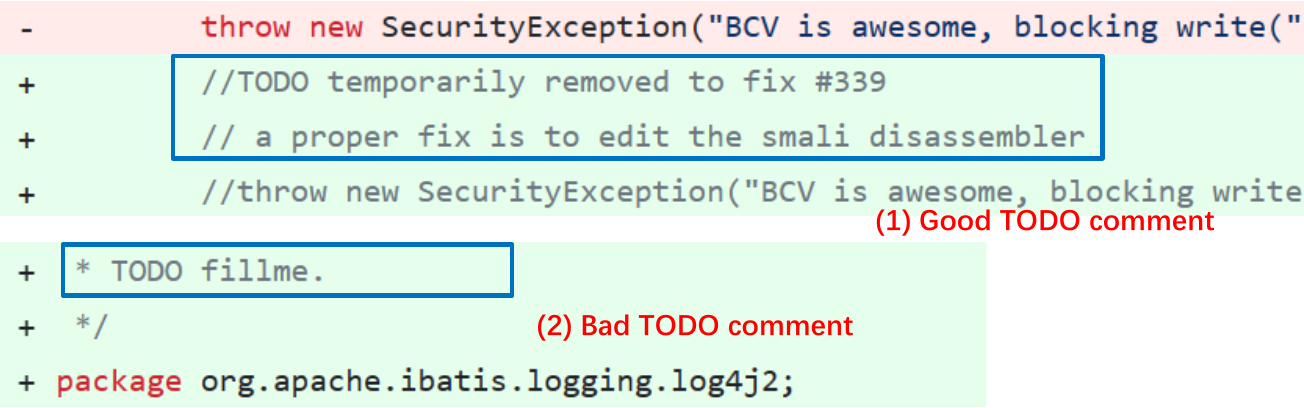}
\caption{Examples of \textit{Good} and \textit{Bad} TODO comments 
}
\label{fig:examples}
\vspace{-0.3cm}
\end{figure}


\textbf{A well-written TODO with sufficient information is vital for code comprehension and software evolution} ~\cite{storey2008todo,ying2005source,nie2018natural}. 
It can prompt the developers for incomplete tasks or remind them to address existing issues effectively. 
Such an example is shown in Fig. \ref{fig:examples} (1). 
It is natural for developers to use such TODOs to record tasks that need to be completed in the short term. However, in practice, there is a large number of TODOs that have existed in software repositories for months or even years. Storey et al.~\cite{storey2008todo} found in their empirical study that `many TODOs remain hidden in the code for years'. Through tracking and analyzing TODOs in a large number of open-source software repositories, we found that the average time from introduction to removal of a TODO is 166.31 days. 
When developers revisit TODOs after a period of time, the quality of the TODOs becomes critically important in facilitating a quick understanding of pending tasks. 

\textbf{Meanwhile, low-quality TODOs can severely impact the development and maintenance of software}~\cite{nie2018natural, sridhara2016automatically, storey2008todo, svensson2015reducing, tan2007icomment, tododie}. 
Such comments can make future changes or improvements to the code more challenging to implement. 
When TODOs lack sufficient information, they introduce ambiguity, hinder the timely completion of pending tasks, and complicate code comprehension and maintenance. 
As TODOs persist in the codebase for extended periods, the negative effects of low-quality TODOs on software development and maintenance become increasingly significant~\cite{haouari2011good, storey2008todo}. 
An example of a ``bad" TODO is shown in Fig. \ref{fig:examples} (2). 
\changed{This TODO is too vague and lacks specific details about the required tasks, making it difficult for developers to recall necessary actions and their reasons.}
In light of the issues with low-quality TODOs, there is a pressing need for further investigation and development in this area.

However, as TODOs are written in natural language it is difficult to test their precision and correctness. Writing high-quality TODOs and maintaining them is a responsibility mostly left to developers. 
In software development, the notation and formatting of TODOs vary greatly. 
Some developers opt to use their own personal style, while others choose to adhere to project-specific conventions when writing TODOs~\cite{storey2008todo, zampetti2021self, cassee2022self}. 
Different TODOs are often embedded with different information (e.g., features, time, and conditions) in semi-structured or unstructured form, making it \textbf{a complex task for ensuring and assessing TODOs quality in practice}. 


So far, most research has focused on TODOs with specific purposes, such as trigger-action TODOs~\cite{nie2019framework} and/or clear goals, such as detecting outdated TODOs~\cite{gao2021automating, sridhara2016automatically}).
Assessing TODO quality has seldom or never been investigated and is still an open research problem. 
First and foremost, \textbf{there is no standard definition of ``good'' or ``bad'' TODOs when it comes to assessing TODOs quality}. 
The absence of a clear definition for TODOs quality not only leads to confusion and miscommunication among developers but also poses significant risks when using existing techniques. 
Recent approaches~\cite{sridhara2016automatically, gao2021automating, nie2019framework, yasmin2022first} collect various TODOs datasets from OSS Projects for training and analysis. 
Unfortunately, these datasets could involve considerable proportions of poorly written TODOs. 
This results in most automated tools artificially inflating the precision of validation experiments and/or drawing invalid conclusions. 
\textbf{There is currently no quality assessment tool available to measure TODOs quality}. 
This further complicates the process of evaluating and improving the quality of TODOs. 
Therefore, it is beneficial for developers as well as researchers to have a definition of what constitutes a ``good'' TODO and a tool to automatically identify good TODOs. 


To address this gap we first collected Top100 popular Java repositories (ranked by stars) on GitHub.
We then manually examined approximately 2,800 TODOs extracted from these repositories.
\changed{We summarized the different types of TODOs and defined and validated a series of criteria as high-quality TODOs based on their types.
The evaluation considers not just writing, but also their clarity and utility to developers.}
Our analysis revealed that roughly 46.7\% of TODOs in open-source software projects are of relatively low-quality, underscoring the negative impact that such comments can have on both software repositories and research efforts.
Next, we conducted a qualitative analysis of 600 high-quality TODOs to better understand the characteristics of well-crafted annotations.
We investigated the introduction and elimination of TODOs of varying qualities, revealing practical implications and inspiring new perspectives on this research topic. 
Finally, to aid developers in proactively recognizing poor quality TODOs, we developed a deep learning-based tool to automate the recognition of well-written TODOs, which demonstrated strong performance in our experiments.

This work makes the following key contributions:
\begin{itemize} 
    \item \changed{A set of novel criteria is proposed for evaluating the quality of TODOs, customized for task based TODOs and notice based TODOs. This criteria provides a structured framework for assessing TODO quality across different contexts; }
    \item \changed{Through an in-depth analysis of TODOs of varying quality, we proposed a taxonomy that links TODOs to nine developer activity types and four purpose categories.}
    \item We manually labeled 2,863 TODOs by our quality criteria and created the first large dataset regarding TODO quality; 
    \item We developed a high performing deep learning-based classifier to aid developers in proactively improving TODO quality.
\end{itemize}

As the first attempt at assessing TODO quality, we hope our research facilitates other researchers and practitioners to improve our approaches. We have released our dataset and source code~\cite{release_TODO}.

\changed{The rest of this paper is organized as follows.
Section~\ref{sec:relat} reviews key related work. 
Section~\ref{sec:design} describes the details of our research approach.
Section~\ref{sec:results} presents the results of our study and Section~\ref{sec:impli} provides a discussion of implications for research and practice.
Section~\ref{sec:threats} discusses the threats to the validity of our work, and 
Finally, Section~\ref{sec:conclusion} concludes the paper.}

\section{Related Work}\label{sec:relat}



\subsection{TODOs Management in Software Engineering: }
Despite the fact that TODOs are widely used by developers, the research works focused on managing TODOs are still limited. 
Empirical studies conducted by Storey et al.~\cite{storey2008todo} found that the use of TODOs varies from individuals to teams, and the incorrect way of managing TODOs can lead to software maintenance issues. 
Nie et al.~\cite{nie2018natural} proposed several techniques for comment and program analysis to support TODOs as software evolves. 
After that, they presented a framework called TrigIt~\cite{nie2019framework}, designed to specify trigger action TODOs in executable format. These actions are then automatically executed when their corresponding triggers evaluate to true.  
Sridhara et al.~\cite{sridhara2016automatically} proposed a rule-based method for identifying outdated TODO comments. 
Following that, Gao et al.~\cite{gao2021automating} proposed a neural-network based model, named TDCleaner, to remove obsolete TODOs by mining histories of software repositories. 
Mohayeji et al.~\cite{mohayeji2022adoption} investigated the impact of the TODO Bot on software development practice by analyzing 2,280 repositories on GitHub.
Yasmin et al.~\cite{yasmin2022first} collected comments tagged with ``TODO'', ``FIXME'', or ``XXX'' from  five popular Apache open-source software projects. 
They investigated the existence and characteristics of duplicate and near-duplicate SATD comments by mining the commit history of a software project.

\changed{In order to facilitate the management of TODOs, a series of tools have also been developed.
Innobuilt Software developed an online tool, named imdone~\cite{imdone}, to extract and track TODOs by creating and updating issues on GiHub or JIRA\footnote{https://www.atlassian.com/software/jira}. 
Similarly, \texttt{todo\_or\_die}~\cite{tododie} is an online tool for keeping TODOs up-to-date by assigning a date and breaking upon outdated TODOs executions. 
TODO Bot~\cite{todobot} is a GitHub application that automates the process of converting TODOs into issues. These tools indicate the industrial focus on refining TODO management.}

Previous studies thus confirm that TODOs play an important role in communication among developers, but no prior work has investigated what is a good TODO comment and how to craft a good one. 

\subsection{Self-admitted Technical Debt in Software Engineering: }
\changed{
Technical debt (TD) occurs when developers opt for suboptimal solutions to achieve short-term goals, potentially compromising the long-term quality of software~\cite{li2015systematic}.
Code comments with TODO, FIXME or XXX tags are often used to represent instances of Technical debt (TD).
Building on the concept of TD, Self-Admitted Technical Debt (SATD)~\cite{potdar2014exploratory} is a subset where developers intentionally introduce suboptimal code implementations and document by code comments.
Many scholars have conducted research on the impact of SATD on software development~\cite{kamei2016using, wehaibi2016examining, russo2022weaksatd} and SATD practice~\cite{cassee2022self, zampetti2021self, fucci2020self, xavier2020beyond}. For example, Wehaibi et al.~\cite{wehaibi2016examining} observed that although SATD typically results in more complicated code changes, changes with SATD tend to yield fewer defects in the future compared to changes without SATD. This finding highlights the significance of identifying and addressing SATD to reduce the occurrence of code defects. Russo et al.~\cite{russo2022weaksatd} pointed out that the presence of SATD may affect the security of software.
To study where SATD will be introduced, Fucci et al.~\cite{fucci2020self} analyzed 5 Java open-source projects and found that SATD was mainly introduced through code changes and code reviews.
}

\changed{In order to identify TD , many prior works have attempted to utilize various data sources. For example, Zazworka et al.~\cite{zazworka2013case} and Nord et al.~\cite{nord2012search} identify SATD through source code. More studies are focused on identifying SATD through code comments~\cite{maldonado2015detecting, huang2018identifying, kamei2016using, ren2019neural}.
For example, Huang et al.~\cite{huang2018identifying} utilized feature selection to select useful features to build classifiers that could identify SATD comments in target projects. Ren et al.~\cite{ren2019neural} build a CNN-based model to determine whether a comment indicates a SATD. Yan et al.~\cite{yan2020just} proposed to utilize the features about code changes to just-in-time detect SATD.
}

\changed{In addition to these studies, there are also some research dedicated to studying the removal of SATD~\cite{zampetti2018self,zampetti2020automatically, maldonado2017empirical}.
da Silva Maldonado et al.~\cite{maldonado2017empirical} discovered that the original authors often remove a considerable amount of Self-Admitted Technical Debt (SATD). Following an analysis of how SATD removal correlates with code changes, Zampetti et al.\cite{zampetti2018self, zampetti2020automatically} introduced SARDELE, a deep learning classifier designed to suggest one of six strategies for SATD elimination.}

\changed{Existing research has significantly explored the impact, practice, identification and management of SATD. However, our study specifically focuses on the quality of TODO comments, a prevalent form of SATD in software development. We also proposed an automatic classifier designed to assess the quality of TODO comments. This tool could help to prioritize and address TODO comments effectively, thereby enhancing code maintainability and reducing technical debt over time.}

\subsection{Comment Quality in Software Engineering: }
Comments are considered as one of the important artifacts for understanding software systems. 
Previous studies have demonstrated that high-quality comments can support software comprehension, bug detection and software maintenance tasks~\cite{tan2007icomment, dekel2009reading}. 
Assessing code comment quality has gained a lot of attention from researchers recently~\cite{khamis2010automatic, zhou2017analyzing, steidl2013quality}.

Various studies conducted surveys with developers to identify good comment attributes. 
Chen et al.~\cite{chen2009empirical} surveyed 137 developers and highlighted several important quality attributes (i.e., adequacy, complete, traceability, consistency, and trustworthiness).
Similarly, Plosch et al.~\cite{plosch2014value} interviewed 88 practitioners and identified consistency, clarity, accuracy, readability, organization, and understandability as the most important attributes. 
Several works have further proposed techniques to automatically assess code comment quality from different aspects. 
For example, Khamis et al.~\cite{khamis2010automatic} assessed the inline comment quality using a heuristic-based approach. 
Steid et al.~\cite{steidl2013quality} used a machine-learning method to assess the documentation comment quality in terms of four quality attributes, i.e., consistency, coherence, completeness, and usefulness. 
Zhou et al.~\cite{zhou2017analyzing} proposed a heuristic and NLP-based approach to check incomplete and incorrect code comments. 

All of the previous studies focus on the quality of explanation comments~\cite{haouari2011good}, which describes the functionalities of the related code. 
The TODO comment is a different comment type~\cite{ying2005source} which describes pending tasks untouched.
The quality of TODOs has never been investigated. 
Different from assessing the general code comment quality with different quality aspects, our work first investigates the quality of TODO comments and develops a tool to automatically evaluate the quality of TODOs.  

\section{Study Design}\label{sec:design}
This study aims to fill the gap of lack of understanding and assessing TODOs quality in practice.  
To achieve this, we explore the quality of TODOs from their distribution, characteristics, life cycle status and automatic assessment. 
Our study attempts to answer the following key research questions:
\begin{enumerate}
    \item \changed{\textbf{RQ1: What is the prevalence of different quality TODOs in software artefacts?} 
    Despite the widespread use of TODOs in open-source software repositories, overall quality of TODOs across repositories remains unknown.}
    \item \changed{\textbf{RQ2: What are the characteristics of different quality TODOs?} We have summarized a set of criteria to distinguish TODOs of different quality at a high level.
    Further, We want to identify a set of characteristics for high-quality TODOs, and conversely characteristics of low or insufficient quality TODOs. This will guide developers on the key aspects to focus on when composing high-quality TODOs across various scenarios, emphasizing the relevant content for each situation.}
    \item \textbf{RQ3: What disparities arise in the management and maintenance of TODOs with varying quality in practice?} 
    How do TODOs evolve in practice, do higher quality TODOs get actioned more frequently, when do low-quality TODOs occur, and how could TODOs quality be improved? 
    The findings of this research question also prompted us to explore the fourth research question.
    \item \changed{\textbf{RQ4: To what extent is it feasible to automatically detect high-quality TODOs?}} 
    Automatic detection of TODOs would assist practitioners in proactively improving the quality of their TODOs. 
    It would assist researchers in more effectively filtering out low-quality TODOs and mitigating the introduction of such threats.
    Meanwhile, high-quality TODOs would have the potential to make developers more willing to solve them, thereby reducing technical debt and improving code quality.
    Those studies related to TODOs can also utilize our work to mitigate the threats posed by data quality issues.
\end{enumerate}

\begin{figure}[]
\centering
\includegraphics[scale=0.6]{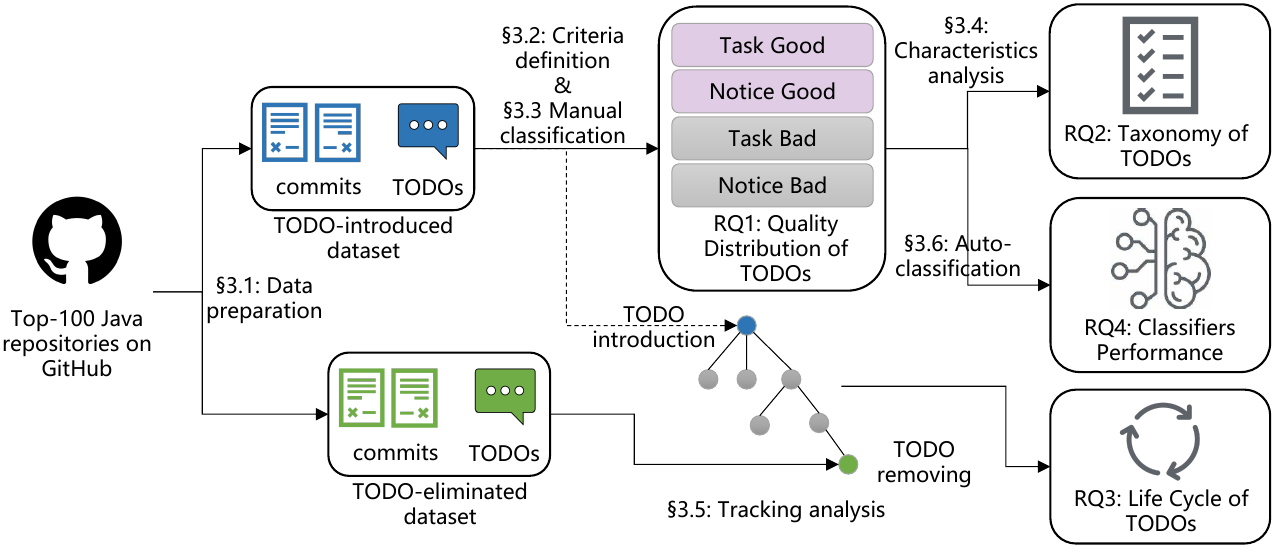}
\caption{Overview of our approach}
\vspace{-0.3cm}
\label{fig:overall}
\end{figure}

We employ a combination of qualitative and quantitative research methods to answer these research questions.
We collected and extracted TODOs from a set of open-source software repositories. 
An overview of our approach is illustrated in Figure~\ref{fig:overall}, which summarizes the overview of our study.
The selection of software repositories, data extraction, and data processing are described in Section~\ref{subsec:data-pre}.
To answer RQ1, we established a set of criteria for defining high-quality TODOs and analyzed the quality of TODOs in open-source repositories based on these criteria.
Our method for identifying good TODOs is described in Sections~\ref{subsec:identify-good} and ~\ref{subsubsec:classify-todo}.
To answer RQ2, we developed a taxonomy to present the characteristics of different types of TODOs, as detailed in Section~\ref{subsec:chara-summ}.
To answer RQ3, we tracked the introduction and removing of TODOs using a specific method, detailed in Section~\ref{subsec:track-method}. 
Finally, to answer RQ4, we developed a deep learning-based model to automatically classify TODOs, as detailed in Section~\ref{subsec:identify-app}.

\subsection{Data Preparation}\label{subsec:data-pre}

To study the criteria and characteristics of high-quality TODOs, we collected a large number of TODOs from recognized popular open-source repositories. 
We assumed that the proportion of high-quality TODOs in these repositories is higher than in less recognized repositories due to the caution exercised by their developers when committing code changes.
Different programming languages may result in language specific content within some code comments, and different programming languages may have different impacts on code quality~\cite{ray2014large, berger2019impact}.
\changed{To avoid these biases, we limited our selection to repositories written in the Java programming language.
According to the Octoverse report of GitHub~\cite{java_top}, Java language ranks among the top 5 most popular programming languages.
Similar to previous studies~\cite{zhang2018multiple, zafar2019towards, tian2022makes}, we started with the Top100 Java repositories from GitHub sorted by their number of stars.} 
Repositories like ``\texttt{LeetCodeAnimation}", ``\texttt{advanced-java}" are then removed because their evolution is different from practical software systems.
These repositories are usually open-source projects for tutorials.

\changed{Some open-source projects do not allow code containing TODOs to be submitted to GitHub~\cite{cassee2022self}, and some developers are unwilling to expose TODOs on open-source platforms~\cite{storey2008todo}.
Hence, some open-source repositories do not include TODOs in the source code published to GitHub, and these repositories also need to be removed.
For each repository, we extracted all the commits from the history up to December 2022.
The corresponding code change (generated by the ``git diff" command), commit message, and commit link for each commit are also extracted. 
Then we iterated through all diffs and used heuristic methods to match and obtain commits involving TODOs.
Once a code change (represented by diff text) includes the keyword ``TODO'' within its code comments, we keep the corresponding commit for that code change.
In this way, the commits containing TODOs can be collected.
The duplicated data pairs of code change and TODO are removed.
In order to reduce the influence of noise data, the non-English and merge commits are all removed.
Finally, the repositories with no remaining commits are all excluded.
After applying these filters, a total of 53 open-source software repositories were retained for our analysis.
}

\changed{We constructed datasets that introduce TODOs and eliminate TODOs based on heuristic rules. 
In our examination of the collected commit diffs from these 53 repositories, if a TODO line starts with a ``+'', we added that commit to the TODO-introduced dataset; conversely, if a TODO line begins with a ``-'', we included that commit in the TODO-eliminated dataset.
}
The TODO-introduced dataset is the main analysis object for RQ1, RQ2 and RQ4. 
\changed{It is also the dataset used to train the classifiers.}
Please note that TODOs in the TODO-introduced dataset are not all TODOs in these repositories.
The TODO-eliminated dataset is used to track the life cycle of TODOs in RQ3.
Finally, we obtain 2,863  \textless{}\texttt{commit}, \texttt{TODO}\textgreater{} pairs in TODO-introduced dataset and 3,313  \textless{}\texttt{commit}, \texttt{TODO}\textgreater{} pairs in TODO-eliminated dataset, respectively.
It can be seen that the TODO-eliminated dataset is bigger than the TODO-introduced dataset.
The reason is that the TODO-eliminated dataset includes many TODOs from branches other than the ``master'' branch. 
\changed{To prevent potential bias, our analysis of the life cycle of TODOs (See Section~\ref{subsec:track-method}) only focuses on determining whether the TODOs within the TODO-introduced dataset are subsequently resolved in commit history.}

\subsection{Criteria for High-quality TODOs}\label{subsec:identify-good}

The quality of TODOs in source code varies greatly, and these TODOs can have varying impacts on repository contributors.
However, there is currently no industry or academic standard for defining high-quality TODOs due to their wide variation in style and content.
Therefore, we established a set of criteria for identifying high-quality TODOs before analyzing their quality, characteristics, and life cycle in open-source software repositories. 
\changed{The criteria were developed based on an exploration of developer forums, a detailed manual analysis of data, and validation by experienced developers. 
The first two detailed procedures are outlined in the following paragraphs, and the details of developer validation are described in section~\ref{subsubsec:classify-todo}.}

In order to obtain insights into what makes high-quality TODOs, we first reviewed papers related to TODO comments and task annotations.
\changed{Firstly, we searched for literature related to ``TODO comments'' on Google Scholar and read articles in the field of software engineering related to TODOs.
Due to the lack of research on TODO quality, we further expanded our scope and instead searched for literature related to technical debt, using a snowball strategy to search for relevant papers as much as possible. We have reviewed a total of 33 papers.}
Unfortunately, most papers do not provide a description of this aspect, but the maintenance of TODOs is a common research topic in the papers.
Therefore we turned to search engines to seek practical perspectives.
\changed{We utilized ``good TODO comment'' and ``TODO comment guidelines'' as initial queries in Google and Bing search engines, and manually checked the Top50 results for each query.
It is important to note that our search process extended beyond these queries, encompassing manual filtering of search results, viewing of the links within the searched web pages, and exploration of additional resources suggested by search engines.
We selected the links that mentioned the quality of TODO comments (links are included in the \changed{replication} package).}
Through qualitative analysis of the content in these web pages, we observed three most commonly mentioned criteria:
1) Use a fixed format (begin with ``TODO"); 
2) \changed{The description is clear enough for developers to understand the context;}
3) Specify an actionable task.
In the website records, we manually checked and the probability of these three criteria being mentioned are 67\%, 76\%, and 48\%, respectively.
Although the first standard is not mentioned the most frequently, it can be said that this standard is a common practice among programmers.
The latter two criteria can be summarized as ``has a clear description of an actionable task".

Many insights on developer forums may reflect an ideal situation. 
In practice, the behavior of developers may deviate from these statements.
A previous study~\cite{nie2018natural} analyzed hundreds of TODOs and found that not all of them contain specific tasks. 
\changed{They} divided TODOs into three categories: task comments, trigger-action comments, and question comments.
Obviously, not all TODOs in open-source repositories specify concrete tasks, and there are also some TODOs that are concerned about potential issues and optional implementation. 
In order to include these criteria summarized above, \textbf{we borrow their insights~\cite{nie2018natural} and classify the form of TODOs into two categories: ``Task TODO" and ``Notice TODO". 
If a task needs to be done is described in the TODO, it is a Task TODO. 
Otherwise, the TODO is labeled as Notice TODO.}
We randomly sampled 200 \textless{}\texttt{commit}, \texttt{TODO}\textgreater{} data pairs from the TODO-introduced dataset (described in Section~\ref{subsec:data-pre}). 
The first two authors of this study carefully read all the samples and summarized the criteria for identifying high-quality TODOs. 
\changed{Based on our preliminary investigation and the personal understanding}, the first two authors labeled the 200 samples independently, categorizing each into one of the four types (i.e., ``Task Good", ``Task Bad", ``Notice Good", ``Notice Bad").
Then the third author reviewed the different parts of labels and discussed with the first two authors to reach a consensus.  Through this approach, the scope boundaries among categories can be clarified.
{Finally, we define two criteria for identifying high-quality TODOs:}
\begin{itemize}
    \item Among Task TODOs, a high-quality TODO should clearly describe a \textbf{specific task} which is \textbf{actionable for stakeholders}. The content about \textbf{``what task''} and \textbf{``how to do''} need to be clear and unambiguous. The task is often written in the form of \textbf{``verb+object''}.
    \item For Notice TODOs, a high-quality TODO should clearly describe \textbf{what is happening}. Then it should \textbf{convey a problem or a potential task} that requires additional attention or even coding from stakeholders.
\end{itemize}

\begin{table}[]
\centering
\caption{Examples for different quality of TODOs}
\label{tab:criteria}
\footnotesize
\begin{tabular}{@{}ll@{}}
\toprule
Category                     & \multicolumn{1}{c}{Examples}                                                                                                                                                                                         \\ \midrule
\multirow{2}{*}{Task Good}   & \begin{tabular}[c]{@{}l@{}}TODO (cushon): switch to hostJavaToolchain after cl/118829419 makes a blaze release\end{tabular}                                                                                         \\ \cmidrule(l){2-2} 
                             & \begin{tabular}[c]{@{}l@{}}TODO: Explicitly add testing for these versions that validates that starting the node\\ after upgrade fails.\end{tabular}                                                                 \\ \midrule
\multirow{2}{*}{Task Bad}    & TODO (b/159359614): enable                                                                                                                                                                                             \\ \cmidrule(l){2-2} 
                             & TODO: do something else with this anyway                                                                                                                                                                              \\ \midrule
\multirow{2}{*}{Notice Good} & \begin{tabular}[c]{@{}l@{}}TODO: we can only conditionally execute type checks if the cast value is not used\\ later on. if there is a cast then the cast value is going to be used in a method guard.\end{tabular} \\ \cmidrule(l){2-2} 
                             & \begin{tabular}[c]{@{}l@{}}TODO: this assumes that the CWD of the Maven process is the plugin \$\{basedir\},\\ which may not be the case\end{tabular}                                                                 \\ \midrule
\multirow{2}{*}{Notice Bad}  & \begin{tabular}[c]{@{}l@{}}TODO: https://issues.jenkins-ci.org/browse/JENKINS-53788 (JDK11 issue on CI)\end{tabular}                                                                                              \\ \cmidrule(l){2-2} 
                             & TODO: s390 port size(FIXED\_SIZE);                                                                                                                                                                                   \\ \bottomrule
\end{tabular}
\end{table}

\changed{The stakeholders here could be developers themselves or other developers, testers, reviewers, or even project managers.
For example, some TODOs may be assigned to testers for additional testing, to some engineers who need to update documentations according to the code changes, or to reviewers for attention on a particular implementation.}
\textbf{In addition to the above two criteria, some TODOs provide trigger conditions and justifications.}
These contents are considered as sufficient conditions rather than necessary conditions in our evaluation.

\changed{For a clearer understanding, two forms of TODOs with different qualities are listed in Table~\ref{tab:criteria}.
Given a TODO, we first determine whether it explicitly specifies a specific task.
The first example in the table shows a Task TODO with a trigger condition. 
It indicates that the pending task ``\textit{switch to hostJavaToolchain}'' needs to be completed after ``\textit{cl/118829419 makes a blaze release}''.
In some cases, many Task Bad TODOs explicitly indicate the need to perform a task. However, the task itself is very vague, often only represented by one verb. 
For example, consider the third case in the table.
During our manual analysis of 200 samples, 30.8\% of the disagreements pertained to this kind of example. Some TODOs that did not describe a concrete task were initially classified as Notice TODO. 
After discussions to resolve these disagreements, we clarified that Notice TODOs do not specify a task within the content of the TODO.
Developers typically point out the issues or considerations, including descriptions of potential issues, suggestions for performance improvements, or comments during code reviews.
Meanwhile, the information it provides is sufficient for developers to understand the potential task.
For example, ``\textit{TODO: can we avoid CCE here? Can we make the exception message better? See issue \#1551}'' is a TODO from ``\texttt{mockito/mockito}'' (13.9k stars on GitHub).
This TODO indicates that the code snippet here is related to the issue \#1551 and two questions have been raised to provide possible directions for work.
Therefore, the TODO is considered to not clearly indicate what task and how to do it.
Although there is no specific explanation, developers will quickly understand that the potential task is to fix issue \#1551 to avoid \texttt{ClassCastException} or optimize the exception message.
However, the creator of this TODO was not sure if it was feasible.
This example is considered as a Notice Good TODO.
Similarly, the fifth and sixth examples in Table~\ref{tab:criteria} describe some existing problems. 
Although no task was specified, developers are capable of deciding the next step based on their experience.
In addition, as shown in the seventh case, TODOs that only contain an external issue link are considered low-quality.
The use of an external issue tracking system is a good practice in project management, but the external link cannot be used as a standalone TODO.
A high-quality TODO should provide sufficient information for developers to quickly understand the tasks to be completed. 
Moreover, during long-term software maintenance, the content pointed to by external links may be deleted, merged, or moved.
This will make such TODOs unreliable.
}




\subsection{Manual Classification of TODOs}\label{subsubsec:classify-todo}
In total, three programmers (the first three authors of this paper) with more than 6 years industrial experience in Java programming participated in the closed coding procedure. 
The coders manually classified the TODO-introduced dataset, which consists of 2,863 \textless{}\texttt{commit}, \texttt{TODO}\textgreater{} pairs.
The detailed steps are as follows:

\textbf{Step 1. Individual classification.}
The first two authors of this study read the data independently.
Each coder reads the TODOs and related information about commits (i.e., commit message, \textit{diff}).
Each pair is categorized into one of the four TODO types: ``Task Good'', ``Task Bad'', ``Notice Good'', ``Notice Bad''. 
On average, the two authors spent 50.2 hours completing the labeling of 2,863 pairs of data.
After labeling the TODOs, Cohen’s kappa coefficient of agreement~\cite{cohen1960coefficient} between the two authors was 80.28, which indicates a high degree of consistency.

\textbf{Step 2. Discuss and merge conflicts.}
Then we merged conflicts by clarifying the misunderstanding.
After individual classification, the third coder merged the primary results of the first two coders. 
Regarding the TODOs labeled differently, the first two coders explained the reasons for their respective labels. 
If the two cannot reach an agreement, the third coder proceeded with arbitration. All the authors participated in all rounds of the discussions and resolved the raised conflicts on data labeling. 
The discussion lasted for about two hours.
During the labeling process, we found many demonstrative pronouns in TODOs. 
If we can easily infer the specific object or scope referred to from code change, we believe the ``object'' is also clear enough.
Similarly, we treated some common sense content the same way.
\changed{Given the high workload and tight deadlines commonly encountered by developers~\cite{meyer2019today, kuutila2020time}, we believe it is reasonable to adopt the above two considerations that reflect these practical constraints.}

\changed{To further validate the labels, we recruited experienced engineers to gain insights into practitioners' perceptions of TODO quality.
We launched a recruitment post on the internal forum of Zhejiang University.
The post provided the approximate task information, its duration, and specific requirements for participants.
Participants are required to have more than 5 years of industrial experience in Java development and good annotation habits in daily development.
Each participant will be paid 75 Chinese Yuan. The compensation is slightly higher than the average level in the forum, with the aim of better attracting experienced developers. In the end, we successfully recruited 10 engineers.
These participants are employed at various leading technology companies (e.g., Huawei, Alibaba, and NetEase), bringing diverse industrial perspectives and experiences to the research.
They possess an average of 7.1 years of industrial experience in software development using the Java programming language.}
\changed{
Similar to some previous studies~\cite{xu2017answerbot, gao2020generating, xu2021post2vec, gao2023code}, we first randomly selected 50 samples from our prepared dataset for the participants to evaluate. 
This sample size was chosen primarily to avoid participants from impatience.
Otherwise, they may blindly answer some questions and introduce bias into the results.}
\changed{For the participants, we first provided guidance and examples to help them understand their task.
Subsequently, each participant classified each sample into one of four TODO types.
For each commit, participants were able to read the code change, commit message, and the corresponding TODO. However, they were forbidden from discussing with each other or making joint decisions. The purpose is to prevent mutual influence among participants, thereby ensuring the objectivity of the data collected.
}
The participants consumed an average of 28.4 minutes, and their results had an average consistency of 75.6\% with our labels.
This indicates that developers substantially agree with our labels.
Additionally, We further use the Fleiss Kappa~\cite{fleiss1971measuring} to measure the agreement between the 10 developers.
The Fleiss Kappa value of their labeling results was 64.36\%, which indicates substantial agreement.

\subsection{TODO Quality Characteristics}\label{subsec:chara-summ}

After identifying different quality data in different forms of TODOs, we explored the key characteristics of these and provide insights into their optimal composition.
Task TODOs and Notice TODOs have different emphasis on conveying information. 
Task TODO focuses on describing the actions of a task. 
The focus of Notice TODO is to describe the motivation or expected correct state of the task.
Therefore, we selected TODO samples from these two major categories for further investigation.
We randomly sampled 150 TODOs labeled ``Task Good", ``Notice Good", ``Task Bad'', and ``Notice Bad'', respectively.
The classification criteria for these data is based on our two hypotheses described in Section~\ref{subsec:identify-good}, and they have been carefully labeled and manually validated.
\changed{
Similar to a previous study~\cite{tian2022makes}, we attempted to explore the subcategories of TODOs and their characteristics through thematic analysis.
Considering that high-quality TODOs offer more comprehensive information related to pending tasks, we firstly concentrate on performing thematic analysis on high-quality TODOs.}

For each ``Task Good" and ``Notice Good" category, \changed{we conducted a thematic analysis according to the following steps}:
(1) The coders carefully read all sampled TODOs to understand what tasks these developers want to coordinate within them.
\changed{Through their initial review of all TODOs, the coders were able to gain a general overview of the writing manner and a rough distribution of tasks involved in these TODOs.
(2) By induction, the coders generated several initial codes that reflected the observed patterns and variety of tasks identified in these sampled TODOs.}
The coders rechecked every TODO, and if necessary, read the commit that introduced the TODO to obtain more context.
For each TODO, the coders classified it into an initial code. 
(3) Then the coders organized and summarized the existing codes.
At this stage, the coders considered how to combine them to form an overall theme.
(4) After obtaining the initial set of themes, the coders reviewed all the TODOs under each theme again. 
The boundaries of each theme become clear with similar themes \changed{being} merged.


\changed{
For low-quality TODOs, we attempted to classify them based on the themes derived from the analysis of high-quality TODOs. 
When encountering TODOs with ambiguous meanings, we carefully examined the associated code changes and commit messages to gain a better understanding of them. 
For those TODOs that remained unclassified, we proceed to create new themes for them.
Through analysis of low-quality TODOs, we can also gain insights into the development practices associated with low-quality TODOs.}

In our thematic analysis, samples labeled as Task TODO and Notice TODO were assigned to two authors (author 1\&2 for Task TODO, author 1\&3 for Notice TODO), respectively.
After the two coders independently completed the above steps, a  discussion was held to resolve the conflict. 
Then they revisited all the data and adjusted the themes.
This process went through a total of two rounds to determine the final theme.
Through thematic analysis of 600 samples, we are able to understand the approximate task types that developers need to annotate with ``TODO".
It also highlights the key patterns in different types of TODO.

\subsection{TODO Lifecycle Analysis}\label{subsec:track-method}

To answer RQ3, we tracked the life cycle of TODOs with different labels.
In Section~\ref{subsec:data-pre}, we constructed the TODO-introduced dataset and the TODO-eliminated dataset with data in the form of \textless{}$\texttt{commit}$, $\texttt{TODO}$\textgreater{}.
We only investigated the data in the TODO-introduced dataset that were removed in latter evolution.
Given a \textless{}$\texttt{commit}_i$, $\texttt{TODO}_i$\textgreater{} pair in the TODO-introduced dataset, we first matched its corresponding removed pair (referred as \textless{}$\texttt{commit}_e$, $\texttt{TODO}_e$\textgreater{}) from the TODO-eliminated dataset according to the following heuristic rules:
\begin{itemize}
    \item \textbf{Rule1:} The comment of introducing TODO $\texttt{TODO}_i$ and the elimination TODO $\texttt{TODO}_e$ should be the same;
    \item \textbf{Rule2:} $\texttt{TODO}_i$ and $\texttt{TODO}_e$ are from the same project;
    \item \textbf{Rule3:} \changed{$\texttt{TODO}_i$ and $\texttt{TODO}_e$ are from a file with the same name;}
    \item \textbf{Rule4:} $\texttt{commit}_i$ is committed earlier than $\texttt{commit}_e$;
\end{itemize}

The reason for the fourth rule is that some revert commits cause the removal of TODO happening earlier than the introduction of TODO.
We cannot consider such TODOs as being addressed.
Please note that if the file where a TODO is located has been renamed, our algorithm can not determine whether the TODO in $\texttt{commit}_i$ and $\texttt{commit}_e$ is the same one.

For every data pair in the TODO-introduced dataset, we searched \textless{}$\texttt{commit}_e$, $\texttt{TODO}_e$\textgreater{} from the TODO-eliminated dataset to see if it can be matched using the above four rules. 
Once $\texttt{TODO}_i$ is matched with $\texttt{TODO}_e$, it means the $\texttt{TODO}_i$ introduced in $\texttt{commit}_i$ is removed in the subsequent reversion $\texttt{commit}_e$. 
We collected all these \textbf{removed TODOs} for later analysis. 
Once a TODO is removed, we consider that it has been resolved in the corresponding commit or has been resolved before, becoming an obsolete TODO and being cleaned up.
However, this assumption is not always true in practice, a number of TODOs are removed for cleanup purposes without addressing them properly. 
In order to further eliminate bias, the first two authors manually checked each removed TODO to see if it is genuinely addressed or removed for clean-up purposes. 
The code change and commit message of each $\texttt{commit}_e$, and the source code of the corresponding version when removing TODO were inspected by the two authors respectively.
Then, further discussion was conducted regarding the different opinions of the two individuals to reach an agreement.
\changed{Additionally, when multiple identical TODOs appear in one file, our matching rules can cause mismatch. 
If only a portion of TODOs are removed, incorrect matching will result in errors. Therefore, we also check for the occurrence of this situation during manual inspection. We found that 10.7\% of the $\texttt{commit}_e$ have duplicate TODOs. But all these TODOs were removed simultaneously in the commit history.
For example, ``\textit{TODO: remove try/catch block when compacting encrypted Realms is supported}'' is a TODO from “realm/realm-java" (11.4k stars on GitHub). This TODO appeared twice within the same file.  The developer removed both TODOs in a single commit with a commit message stating, ``\textit{Removing the restriction of compacting encrypted Realm files.}'' 
After lifting the restriction on compacting encrypted Realm files, the developer deleted these TODOs altogether.
Therefore, we only record it as one removal when we count the number of removed TODOs. As for the other statistical data we will soon mention later, this situation has no impact.}

\changed{Finally, all the removed TODOs are classified into two groups: \textbf{unresolved TODOs} (TODOs that are removed without addressing them) and  \textbf{resolved TODOs} (TODOs that are removed because being addressed). 
The life cycle information of the TODOs were parsed, including:}
\changed{
\begin{enumerate}
    \item \textbf{Resolved Proportion.} \texttt{Resolved\%} $=($the number of resolved TODOs $/$ the number of all TODOs of this type$)$; 
    \item \textbf{Unresolved Proportion.} \texttt{Unresolved\%} $=($the number of unresolved TODOs $/$the number of removed TODOs$)$;
    \item \textbf{Time-interval between introduction and resolved.} \texttt{Time-Interval} $=(\texttt{commitTime}_e - \texttt{commitTime}_i)$; 
    \item \textbf{The number of commits experienced}, \#\texttt{Commits} $ = \texttt{DFS}$ $(\texttt{commitNode}_e, \texttt{commitNode}_i)$, where $\texttt{DFS}$ is a Depth-first search algorithm, $\texttt{commitNode}_e$ and $\texttt{commitNode}_i$ are the commit node in the commit history (as a tree structure).
\end{enumerate}}

\subsection{Automatic TODO Classification}\label{subsec:identify-app}

A well written TODO can help developers quickly understand the tasks they need to complete and the points they need to pay attention to.
This helps reduce the time it takes for developers to recall the context when checking legacy TODOs, thus improving development efficiency.
Further, according to the results of RQ3 (Please refer to Section~\ref{subsec:lifecycle-result} for details), the developers are more willing to first address high-quality TODOs.
Therefore, we hope to build a tool to help identify high-quality TODOs. 
This will potentially improve the quality of TODOs written by developers and also save developers' effort by prioritizing which TODOs to address first. 
The existence time and impact of technical debt in software repositories would be reduced, resulting in improved code quality.
In addition, for research related to TODOs, this tool can help mitigate the bias caused by the quality of TODOs.

Finally, in order to answer RQ4, we developed a deep learning-based approach to automatically identify high-quality TODOs.
When defining the criteria for identifying high-quality TODOs, we found that TODOs in our examined open-source software repositories can mainly be divided into two forms, e.g., Task TODO and Notice TODO.
Therefore, we built two classifiers. 
The first classifier is to determine whether a TODO is a Task TODO or a Notice TODO.  
A second classifier is applied to identify high-quality TODOs of each type.

\textbf{Experimental Data: }
In the previous stages, the TODO-introduced dataset has been carefully labeled and validated by experienced programmers.
Therefore, we chose this dataset to train and test our classifiers.
Due to the presence of some noisy tokens in TODOs, such as issue id, creator's information, etc. 
Therefore, we further cleaned TODO comments to avoid the influence from these noises.
We first merged multiple lines of TODOs to form a complete paragraph.
Some programmers may mark their identity or contact information at the beginning of the TODOs, while others record the TODO related external link there.
For example, ``\textit{TODO(b/20335397): This code was relying on Bitmap equality which Robolectric removed}".
Therefore, we replaced the content in parentheses with ``\textless{}\texttt{info\_tag}\textgreater{}".
We also replaced some TODO content with placeholders. 
Specifically, we replaced commit id with ``\textless{}\texttt{commit\_id}\textgreater{}",  and replaced issue id or pull request id like ``\texttt{\#3072}'' with ``\textless{}\texttt{link\_id}\textgreater{}".
Finally, the processed TODOs are tokenized with white spaces and punctuation.


In order to make the classifier better learn the correlation between TODOs and code implementations, we use the code changes as one of the corpus for classifier learning.
We carefully followed the cleaning steps for code changes in the study from Gao et al.~\cite{gao2021automating}.
The code changes are characterized by {\it diff}, which can be generated by the {\it git} {\it diff} command in Git.
The {\it diff} header was deleted and the commit id was replaced with ``\textless{}\texttt{commit\_id}\textgreater{}".
Then we converted the {\it diff} into lowercase and tokenized it with spaces and punctuation.
In this way, we obtained the cleaned TODOs and their corresponding code changes for model training and inference.

\begin{figure}[]
\centering
\includegraphics[scale=0.6]{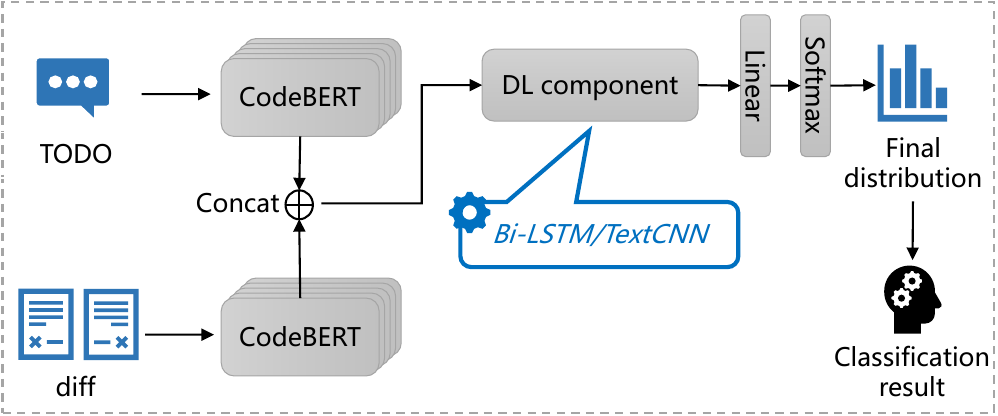}
\caption{Structure of deep learning-based classifiers}
\label{fig:approach}
\end{figure}

\textbf{TODO Classifiers: }
At present, many effective methods have been proposed for code document classification~\cite{yang2019survey, tian2022makes, li2024empirically}.
Our purpose is to explore the possibility of high-quality automatic identification, so we constructed classifiers based on deep learning algorithms.
We selected some machine learning based methods and deep learning based methods with CodeBERT~\cite{feng2020codebert} to construct our classifiers.
Some previous studies~\cite{feng2020codebert, lin2021traceability, rai2022review, di2023code, chai2022cross} have shown the effectiveness of CodeBERT for encoding code documentation.
CodeBERT is a transformer based model which is pre-trained with functions and corresponding natural language descriptions.
It can help capture and represent the semantics of TODOs in classifiers.
\changed{In addition, we also utilized the NLTK library~\cite{loper2002nltk} to construct a baseline based on part-of-speech analysis. This baseline classifies samples by determining whether there is a "verb+object" form in TODOs.}

For machine learning based classifiers, we first used CodeBERT to embed every TODO into a fixed length vector.
We adopted some widely used methods in machine learning, including Multinomial Naive Bayes algorithm~\cite{schutze2008introduction}, Logistic Regression~\cite{tolles2016logistic}, Multi-Layer Perceptron~\cite{riedmiller2014multi}, K-Nearest Neighbors~\cite{peterson2009k}, Random Forest~\cite{breiman2001random}, and Gradient Boosting Machine~\cite{friedman2001greedy}
The vector representations of TODOs are used as input for training and testing of these machine learning based algorithms.
For each input TODO, these machine learning algorithms output a probability distribution and decide the classification result.

For deep learning based classifiers, we used two CodeBERT components as the encoder for the TODOs and {\it diffs}, respectively.
\changed{Due to the possible correlation between source code, code comments in {\it diffs} and TODOs, we also consider {\it diffs} as one of the inputs during model construction like some previous studies~\cite{liu2018automatic, liu2020automating, gao2021automating}.}
The output matrix of the two encoders then are concatenated as a separate matrix.
This concatenated matrix is input as a joint representation of TODO and source code into different components to form different classifiers.
In specific, we constructed a component based on double-layer Bi-LSTM~\cite{schuster1997bidirectional} and a component based on TextCNN~\cite{chen2015convolutional, zhang2015sensitivity}, respectively.
Through this component, our models are able to learn the implicit correlation between TODO and code implementation.
Finally, the output of this component is sequentially passed through a Linear layer and a Softmax layer.
The models determine the classification results based on the probability distribution of the final output.
Figure~\ref{fig:approach} shows the structure of our proposed TODO classifiers.

To ensure fairness in comparison, we use CodeBERT to represent all the input as a vector with a length of 768.
The max length setting of the CodeBERT tokenizer is set to be 512.
We perform Grid Search to ensure reasonable parameter selections as much as possible.
For the deep learning based classifiers, we use AdamW~\cite{loshchilov2018fixing} optimizer algorithm to optimize our network.
All the experiments are conducted on a Ubuntu 20.04 server with two Nvidia GeForce RTX 3090 GPU and 24G memory, and 20 cores 3.7GHz CPU and 32GB memory.

\section{Results}\label{sec:results}

We present the results of our study that answer our four research questions.
Section~\ref{subsec:distribution} discusses the quality distribution of TODOs in the open-source software repositories.
Section~\ref{subsec:result-class-todo} describes our classification of TODOs.
Section~\ref{subsec:lifecycle-result} discusses our analysis of the lifecycle of TODOs.
Finally, 
Section~\ref{subsec:tool} discusses the performance of our constructed machine learning-based classifiers.

\subsection{RQ1: Quality Distribution of TODOs}\label{subsec:distribution}
We manually labeled a TODO-introduced dataset based on the criteria we summarized (Section~\ref{subsubsec:classify-todo}).
This TODO-introduced dataset contains 2,863 pairs of data from 53 popular GitHub repositories.

Firstly, we present a category distribution for all 53 projects.
The final results of our manual classification are presented in Table~\ref{tab:todo-intro-dataset}.
The distribution shows that \textbf{Task TODO accounts for 59.55\% of TODOs}. 
What is not mentioned in opinions from developer forums is that many TODOs do not specify a task to be completed (40.45\% in this dataset). 
These TODOs may be used for communication, problem confirmation, or other purposes.
More details of these categories can be found in Section~\ref{subsec:result-class-todo}.
In addition, we can see that \textbf{for Task TODOs,  low-quality TODOs account for 39.9\%}. In contrast, \textbf{for Notice TODOs, the low-quality TODOs account for 57.3\%}.
This means that there are a large number of TODOs that are considered low-quality, and developers cannot obtain enough information based on TODOs to understand the pending tasks.
Especially when the TODO creator does not state specific tasks, the problem of low-quality TODOs becomes even more severe.
This also indirectly demonstrates the potential threats in current research using TODOs for data analysis and model construction.

\begin{table}[]
\centering
\caption{Counts of Different Categories in the\\ TODO-introduced dataset}
\label{tab:todo-intro-dataset}
\setlength{\tabcolsep}{9mm}{
\begin{tabular}{@{}lcc@{}}
\toprule
\textbf{Category} & \textbf{Count} & \textbf{Proportion} \\ \midrule
Task Good       & 1,024           & 35.77\%             \\
Task Bad          & 681            & 23.78\%             \\
Notice Good       & 495            & 17.29\%             \\
Notice Bad        & 663            & 23.16\%             \\ \bottomrule
\end{tabular}
}
\end{table}

\begin{figure}[]
\centering
\includegraphics[scale=0.6]{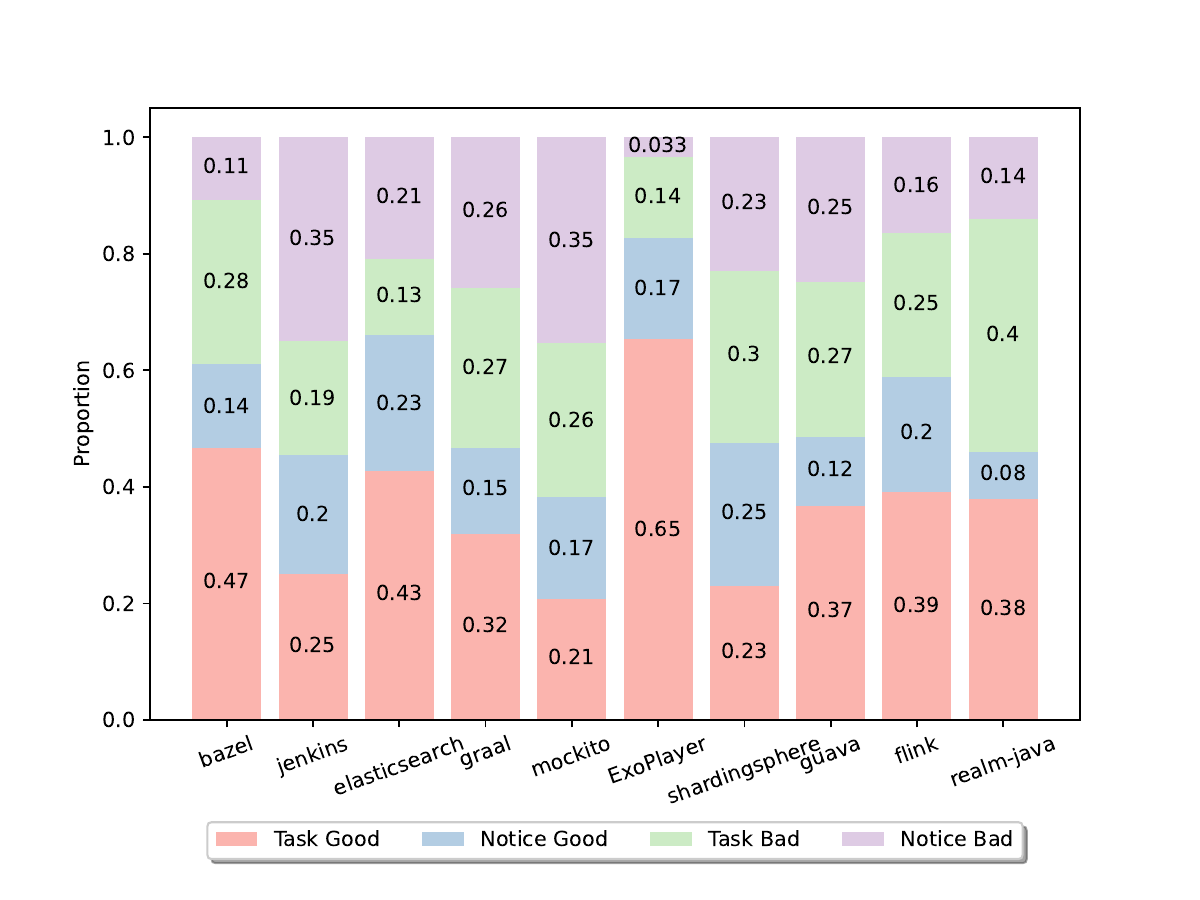}
\vspace{-0.1cm}
\caption{Distribution of different types of TODOs in top-10 projects}
\label{fig:rq1}
\end{figure}

In order to gain better insights into the quality distribution of TODOs based on project granularity, we further selected the projects with the top 10 TODO quantities.
Figure~\ref{fig:rq1} shows the distribution of different types of TODOs in the 10 projects.
We can see that the distribution of these four types of TODOs varies among different open-source software repositories. 
In the ten projects, \textbf{the ratio of Task Good TODOs varies from ca. 21\% to ca. 65\%}, while \textbf{the ratio of Notice Good TODOs varies from ca. 8\% to ca. 25\%}.
In addition, we merged high-quality types of TODOs and recalculated the proportion.
The \textbf{portion of High-quality TODOs in the ten projects varies from ca. 38\% to ca. 82\%}.
This phenomenon indicates that even in these popular software repositories on GitHub, the distribution of TODOs quality varies greatly in project granularity.
Some project contributors may have reached a certain consensus on the quality of the code documentations, so they will pay more attention to the quality of TODOs. 
For example, the proportion of high-quality TODO in the ExoPlayer repository (20.7k stars on GitHub) has reached 82\%.
Some projects may overlook the importance of TODOs, leading to the existence of a large number of low-quality TODOs. 
As shown in Figure~\ref{fig:rq1}, the proportion of high-quality TODOs in mockito repository is only 38\%.
As we mentioned earlier, many scholars utilize information about TODOs and its corresponding source code to build and train their tools.
However, when the quality of TODO is low, they may learn or extract incorrect mapping relationship, a major threat.

\find{
{\bf Summary for RQ1:} 
\changed{Low-quality TODOs constitute a notable portion of open-source software projects.}
Out of the 2,863 TODOs examined in our study, 46.94\% TODOs are of low-quality and need further improvement. 
}

\subsection{RQ2: What makes high-quality TODOs}\label{subsec:result-class-todo}
RQ2 explores the characteristics of different quality TODOs.
After labeling TODOs of different types and qualities, we employed thematic analysis to analyze TODOs across various subcategories.
Specifically, Task TODOs focus on describing specific tasks, and 
Notice TODOs do not explicitly state the action but stakeholders can easily understand potential tasks.
We identified the types of task described in Task TODOs and the purpose of Notice TODOs.
Table~\ref{tab:rq2} shows the overall prevalence of TODOs across various subcategories with different quality levels.
We then presented the various types of Task TODOs and Notice TODOs, respectively.
The detailed characteristics and the optimal composition about each category, and its relevant cases are introduced subsequently.

\begin{table}[]
\centering
\caption{Subdivision Categories of Different Forms of TODOs}
\label{tab:rq2}
\footnotesize
\begin{tabular}{@{}cccc@{}}
\toprule
\multirow{2}{*}{\textbf{TODO}} & \multirow{2}{*}{\textbf{Category}} & \multicolumn{2}{c}{\textbf{Proportion}}      \\ \cmidrule(l){3-4} 
                               &                                    & \textbf{High-quality} & \textbf{Low-quality} \\ \midrule
\multirow{9}{*}{Task TODO}     & Feature Request                    & 30.7\%                &   34.0\%              \\ \cmidrule(l){2-4} 
                               & Remove Workaround                  & 20.0\%                &   16.7\%             \\ \cmidrule(l){2-4} 
                               & Re-enable Commented Code           & 12.0\%                &   10.0\%             \\ \cmidrule(l){2-4} 
                               & Refactoring                        & 11.3\%                &   10.7\%             \\ \cmidrule(l){2-4} 
                               & Bug Fix                            & 9.3\%                 &   12.7\%             \\ \cmidrule(l){2-4} 
                               & Code Hygiene                       & 5.3\%                 &   3.3\%             \\ \cmidrule(l){2-4} 
                               & Problem Confirmation               & 5.3\%                 &   6.7\%                   \\ \cmidrule(l){2-4} 
                               & Testing                            & 4.0\%                 &   2.7\%              \\ \cmidrule(l){2-4} 
                               & Documentation                      & 2.0\%                 &   3.3\%              \\ \midrule
\multirow{6}{*}{Notice TODO}   & Provide suggestion                 & 47.3\%                &   20.0\%                \\ \cmidrule(l){2-4} 
                               & Highlight existing issue           & 25.3\%                &   27.3\%             \\ \cmidrule(l){2-4} 
                               & Analyze current situation          & 16.0\%                &   15.3\%             \\ \cmidrule(l){2-4} 
                               & Throw pending question             & 11.3\%                &   10.7\%             \\ \cmidrule(l){2-4} 
                               & Auto-generated TODOs               &  -                    &   8.7\%             \\ \cmidrule(l){2-4} 
                               & Indecipherable TODOs                    &  -                    &   18.0\%              \\ \bottomrule
\end{tabular}
\end{table}

\subsubsection{Task TODOs}
We identified nine categories of pending task described by Task TODOs.
These task types basically cover the activities that programmers will engage in during the development and maintenance process.
This also demonstrates the widespread use of TODOs in program development, highlighting their importance.
\begin{enumerate}
    \item \textbf{Feature Request:} This subcategory requests the implementation of a new feature, function, or program logic, and is the most common Task TODOs type. For example, \textit{``TODO support re-consumable for HYBRID\_FULL resultPartitionType.''}. \changed{Many low-quality TODOs of this subcategory only contain one verb. For example, \textit{``TODO implement''} and \textit{``TODO add''}.}
    \item \textbf{Remove Workaround:} These Task TODOs mean that the relevant code change is a temporary solution introduced for a certain goal like defect repair and performance improvement.  Therefore, developers need to replace the workaround with a formal solution in the future. 
    The scope of the workaround and triggering conditions are clearly defined. For example,\textit{ ``TODO(b/30281236): Remove the flag after deprecation.''}. \changed{However, the low-quality TODOs of this task type often do not accurately indicate which relevant code snippets need to be removed, such as \textit{``TODO remove me''} and \textit{``TODO remove this''}.}
    \item \textbf{Re-enable Commented Code:} Task TODOs of this subcategory are followed by a commented code fragment. This implementation may be temporarily commented due to issues such as dependency packages, version matching, and feature support. For this subcategory, the most important thing is the commented code that needs to be restored and the trigger conditions for restoration. For example, \textit{``TODO(fry): enable this check once the default is changed to NULL\_STATS\_COUNTER''}. \changed{Similarly, this subcategory of low-quality TODOs often fail to convey sufficient information. For example, \textit{``TODO uncomment''} and \textit{``TODO reenable after the upgrade''}.}
    \item  \textbf{Refactoring:} These Task TODOs request the improvement or optimization of existing code implementation. These TODOs clearly describe the object and purpose of the refactoring. Some developers may provide alternative solutions. For example, \textit{``TODO: Refactor this function to reduce its complexity''}.
    \changed{The low-quality TODOs of this subcategory often miss the information about the purpose. For example,  \textit{``TODO: Do something better with this case''}.}
    \item \textbf{Bug Fix:} These Task TODOs identify a bug or issue in the code and request its resolution. Developers often indicate what the problem is, analyze the reasons or suggest solutions. If there are any,  issue links or external reference materials are provided. For example, \textit{``TODO(simonw): fix this method to select the oldest del gen if we pick a del file''}. \changed{Many low-quality TODOs of this subcategory only write "TODO: fix that" or "TODO fix me" without providing any other information. So when they are left in the code repository for a period of time, developers will find it difficult to recall the context again.}
    \item  \textbf{Code Hygiene:} This subcategory requests programmers to perform clean-up tasks unrelated to the logic of code, such as removing comments, moving code, and removing obsolete code. For example, \textit{``TODO Seems unused. Delete when confirmed.''}. \changed{This type of low-quality TODOs usually only mention cleaning up. But whether to directly delete the code or move it to another location is not specified.}
    \item \textbf{Problem Confirmation:} This subcategory requests developers to confirm some important issues, such as security risks, code logic vulnerabilities, or the root cause of the defect. For example, \textit{``TODO confirm safe to assume non-null and use getInstance()''}. \changed{This type of low-quality TODOs usually do not specify which aspect of the problem needs to be checked or confirmed, such as \textit{``TODO: look at this one later''}.}
    \item \textbf{Testing:} This subcategory requests the addition or update of tests for a code block or module. The TODOs sometimes are with a trigger condition (such as when a function is completed or a defect is fixed) to request developers to add corresponding test code. For example, \textit{``TODO(anuraaga): Add unit tests after https://github.com/line/armeria/issues/2220''}. \changed{Similar to low-quality TODOs in other subcategories, many TODOs simply state \textit{``TODO: test me''}. However, the context does not provide clarity on the specific aspects to be tested.}
    \item \textbf{Documentation:} This subcategory requests the creation or update of documentation for a code snippet, a function, or a class. The proportion of tasks in this subcategory is relatively low. For example, \textit{``TODO (thomaswue): Document why this must not be called on floating nodes.''}. \changed{Many low-quality TODOs of this type only contain phrases like \textit{``TODO fill me''} or \textit{``TODO finish this''}, which can easily lead to confusion with other subcategories. Typically, a review of the code is required to understand that these TODOs are prompting for documentary work.}
\end{enumerate}

\subsubsection{Notice TODOs}
\changed{For the Notice TODOs, we identified four categories of purposes for introducing these TODOs and two categories that only appear only in low-quality Notice TODOs.
Since these TODOs do not directly state what to do, they are often written in a way that describes what is happening or raises questions.
For the additional two subcategories, since they do not convey any intent from the code submitter, we do not include them in the categories of purposes.}
Details about these categories with examples are as follows:
\begin{enumerate}
    \item \textbf{Provide suggestion:} this subcategory provides some suggestions for pending tasks, such as code implementation ideas, optimization solutions, and defect solutions. This type of TODOs may provide a possible task, but it requires further evaluation by developers. This subcategory has the highest frequency of occurrence, 47.3\% of Notice TODOs. For example, \textit{``TODO better to use ModelHyperlinkNote.encodeTo(User), or User.getUrl, since it handles URL escaping''}. \changed{Many low-quality TODOs related to this purpose only offer a vague comment or a code snippet, making it challenging to obtain specific and valuable suggestions directly from the TODOs themselves. For example, \textit{``TODO cf.ensureSuccess();''}.}
    \item \textbf{Highlight existing issue:} this subcategory is mainly for highlighting existing issues. In addition, this subcategory analyzes the reasons for the issue, and, if possible, reference materials related to the issue could be attached in the form of links. For example, \textit{``TODO(janakr): this is failing since the test was disabled and someone snuck a regression in. Fix.''}. \changed{Most of these low-quality TODOs typically only contain an external link or an issue ID. However, these TODOs that point to external information may become unreliable due to the movement, merging, or deletion of linked content.}
    \item \textbf{Analyze current situation:} this subcategory raises concerns about potential defects by analyzing the current code implementation and program operation. These TODOs sometimes describe the possible consequences of not taking action. The developers of these TODOs determine the subsequent tasks to be completed based on the information provided by TODOs. For examples, \textit{``TODO: FullyQualifiedName.Factory\#create has an overload which accepts `package'.''}. \changed{Many of these low-quality TODOs express useless or unclear comments about the current situation. For example, \textit{``@TODO This really shouldn't be happening''}.}
    \item \textbf{Throw pending question:} this subcategory requests developers to further confirm the concerns raised by the creator through the form of a question. These Notice TODOs may be a challenge about the current implementation or a vision for future implementations. For example, \textit{``TODO we would need a .type property on reducers too for this error message?''}. \changed{The low-quality TODOs of this subcategory may only raise a brief question without providing additional information about concerns. Sometimes, it's merely a line of code followed by a question mark. For example, \textit{``TODO(will): herm... doesn't look quite righT?''}.}
    \item \changed{\textbf{Auto-generated TODOs:} Auto-generated TODOs refer to TODOs that are automatically inserted into the code by development tools, frameworks, or during code generation processes. Unlike manually added TODOs, which reflect developers' immediate thoughts or intentions, auto-generated TODOs often have a standard template or format. The specificity of Auto-generated TODOs is poor, which may lead to the accumulation of debt as the project progresses. For example, \textit{``TODO Auto-generated catch block''} and \textit{``TODO Auto-generated method stub''}.}
    \item \changed{\textbf{Indecipherable TODOs:} this subcategory includes TODOs that are either incomprehensible or meaningless. These TODOs typically only contain fragments or a few simple symbols, making it challenging for developers to understand the intended task or the specific issue that needs to be addressed. These low-quality TODOs pose a challenge in software maintenance, as they can not effectively convey their purpose or requirements to the stakeholders. For example, \textit{``TODO ...''} and \textit{``TODO ewwww''}.}
\end{enumerate}

\find{
{\bf Summary for RQ2:} We identified nine task types and four purpose types for Task TODOs and Notice TODOs respectively, along with their respective expression preferences. 
These criteria could serve as valuable inspiration for developing TODO-related generative tools and assist developers in writing high-quality TODOs.
}

\subsection{RQ3: Life Cycle of TODOs}\label{subsec:lifecycle-result}

TODOs are introduced by developers to denote the pending tasks. 
However, many TODOs are left in the code and never get revisited after their introduction. 
In this research question, we investigated the life cycle of TODOs regarding their varying qualities. 
Particularly, we investigated if high-quality TODOs get resolved more regularly than low-quality TODOs in software development practice. 
\textbf{Since not all removed TODOs are guaranteed to be addressed (some TODOs are removed for no reasons or just clean-up the codebase), for each removed TODO commit, we manually label the TODO as genuinely addressed or clean-up purposes by investigating their associated code change and commit messages.}
\changed{Table~\ref{tab:todo-lifecycle} shows the lifecycle of Task TODOs and Notice TODOs, in terms of resolved proportion, unresolved proportion (the ratio of TODOs that have been removed but not resolved),  time interval (the average time cost for addressing TODOs on the \textbf{resolved TODOs}) and the number of commits (the average number of commits between TODOs introduction and resolution on the \textbf{resolved TODOs}). }

\changed{The main focus of our investigation is the differences between high-quality TODOs and low-quality TODOs. 
Therefore, we further leveraged Wilcoxon rank-sum test~\cite{wilcoxon1992individual} with a Holm-Bonferroni correction~\cite{holm1979simple} to analyze statistical significance of difference in means of these feature values between high-quality TODOs and low-quality TODOs respectively. 
We tested the following hypotheses:
\begin{itemize}
    \item  \textbf{Null Hypothesis} (\(H_{0,1}\)): There is no significant difference in the proportion of resolved high-quality TODOs compared to resolved low-quality TODOs. \\
    \textbf{Alternative Hypothesis} (\(H_{A,1}\)): The proportion of resolved high-quality TODOs significantly differs from that of resolved low-quality TODOs.
    \item \textbf{Null Hypothesis} (\(H_{0,2}\)): There is no significant difference in the proportion of unresolved TODOs between high-quality removed TODOs and low-quality removed TODOs. \\
    \textbf{Alternative Hypothesis} (\(H_{A,2}\)): There is a significant difference in the proportion of unresolved TODOs between high-quality removed TODOs and low-quality removed TODOs.
    \item \textbf{Null Hypothesis} (\(H_{0,3}\)): There is no significant difference in the time interval between the introduction and resolution of high-quality resolved TODOs compared to low-quality resolved TODOs. \\
    \textbf{Alternative Hypothesis} (\(H_{A,3}\)): The time interval between the introduction and resolution of high-quality resolved TODOs significantly differs from that of low-quality resolved TODOs.
    \item \textbf{Null Hypothesis} (\(H_{0,4}\)): There is no significant difference in the number of commits experienced by high-quality resolved TODOs compared to low-quality resolved TODOs. \\
    \textbf{Alternative Hypothesis} (\(H_{A,4}\)): The number of commits experienced by high-quality resolved TODOs significantly differs from that of low-quality resolved TODOs.
\end{itemize}
}

\changed{
The independent variables of the tests are different quality of TODOs (i.e., High-quality vs. Low-quality).
The dependent variables are the resolved proportion of different projects, the unresolved proportion of different projects, the time interval of different resolved TODOs, and the number of commits of different resolved TODOs.
}

\begin{table}[]
\centering
\caption{Lifecycle of Different TODO Categories in the TODO-introduced dataset}
\label{tab:todo-lifecycle}
\footnotesize
\begin{tabular}{@{}lcccc@{}}
\toprule
Category & Resolved\% & Unresolved\% & Time-Interval & \#Commits \\ \midrule
Task Good       & 17.97\%  & 5.15\% & 119.68 (day) & 600.80 \\
Notice Good     & 10.91\%  & 8.47\%  & 77.43 (day) & 447.61  \\
\textbf{High-quality}    & \textbf{15.67\%}  & \textbf{5.93\%} & \textbf{98.56 (day)} & \textbf{524.21}  \\ \hline 
Task Bad        & 15.57\%  & 19.70\% & 270.79 (day) & 2152.44  \\
Notice Bad      & 10.70\%  & 39.83\% & 198.80 (day) & 644.72  \\
\textbf{Low-quality}     & \textbf{13.24\%}  & \textbf{29.20\%} &  \textbf{234.79 (day)}   & \textbf{1398.58}  \\\hline 
Overall      & 14.50\%  & 17.50\% & 166.31 (day) & 984.71  \\ 
\bottomrule
\end{tabular}
\end{table}

\changed{
From Table~\ref{tab:todo-lifecycle}, we can see that (i) the proportion of high-quality TODOs being addressed is only slightly higher than that of low-quality TODOs.
In general, only a small amount (less than 20\%) of TODOs are removed for addressing them.
This phenomenon further reflects the current situation of TODOs in many software repositories, with a large number of TODOs being forgotten in the source code after being created.
(ii) According to the Unresolved ratio, developers tend to clean up low-quality TODOs in software practice. Even though these TODOs have not been truly resolved.
For example, around 40\% of Notice Bad TODOs are removed without addressing them, while the ratio is much lower (i.e., 8.47\%) for Notice Good TODOs. 
This phenomenon is understandable, considering that lower-quality TODOs typically exert a diminished impact on software quality and maintenance.
In practice scenarios, when developers come across TODOs, the low quality TODOs are more likely to be disregarded or removed due to their lack of effective information about the tasks to be completed.
In contrast, high-quality TODOs, characterized by their clarity and provision of actionable instructions, are often taken more seriously by developers.
(iii) Low-quality TODOs have longer time-interval and larger number of commits than high-quality TODOs -- indicating that low-quality TODOs do not provide clear and actionable information, leading to significant delays and/or efforts for addressing them. 
For instance, developers spend 119 days on average to address a Task Good TODO, while the time cost for addressing a Task Bad TODO is more than doubled (270 days). 
Task Good TODOs goes through an average of 600.8 code submissions from being introduced to being resolved. Task Bad TODOs experiences an average of 2152.44 commits, which is three times more than Task Good TODOs.
The similar situation also goes with Notice Good and Notice Bad TODOs. 
These contrasts clearly show how the quality of TODOs significantly affects developer decisions and directly impacts the efficiency of software maintenance and development.
}

\changed{
As stated above, we conducted Wilcoxon rank-sum tests with Holm-Bonferroni corrections to test our hypotheses.
The initial significance levels were preset at 0.05.
We compared the p-values obtained from the Wilcoxon rank-sum test against the adjusted significance levels for each hypothesis.
Our results indicate that the second and third hypotheses were supported by the data, leading to the acceptance of the alternative hypotheses.
While the first and fourth hypotheses did not show statistical significance, resulting in the acceptance of the null hypotheses.
These results indicate a significant difference in the proportion of unresolved TODOs between high-quality and low-quality removed TODOs. 
Additionally, the time interval between the introduction and resolution of high-quality resolved TODOs significantly differs from that of low-quality resolved TODOs. 
However, when examining the overall proportion of TODOs that were resolved, as well as the number of commits in the life cycle of TODOs, no significant difference was observed between low-quality and high-quality resolved TODOs.
}

\find{
{\bf Summary for RQ3:}
Developers are more inclined to resolve high-quality TODOs, which cost significantly less time to address as compared with low-quality TODOs.  
In addition, low-quality TODOs are more likely to be cleaned indiscriminately by developers. 
}

\subsection{RQ4: Automatically Classifying TODOs}\label{subsec:tool}

Following on from the answers to RQ1-RQ3, there is a need for developers to be able to write high quality TODOs and proactively improve low quality TODOs. We explore in RQ4 how well we might automatically classify TODO quality in practice to support this. 
\changed{As described in Section~\ref{subsec:identify-app}, we constructed one method based on part-of-speech analysis, and 8 methods based on machine learning algorithms and deep learning algorithms for classifying TODOs.
}
For machine learning based classifiers, we directly embedded TODOs using the pre-trained model CodeBERT, and then fed it to the machine learning algorithm.
But for deep learning based classifiers, we further used our training data for fine-tuning.
To reduce variance, we performed a 10-fold cross-validation on labeled TODO-introduced dataset to evaluate the effectiveness of each classifier.
Meanwhile, in order to make a fair comparison, we fine-tuned the model's hyperparameters (e.g., n\_estimators in Random Forest, batch size, learning rate of deep learning models) and chose the one with the best performance. We report the average scores of the 10 folds for all the nine methods in Table~\ref{tab:rq3}.

\changed{The Part-of-Speech baseline did not learn patterns from training dataset like machine learning based methods.
From Table~\ref{tab:rq3}, it can be seen that the performance of Part-of-Speech baseline is similar to some machine learning based methods.
This phenomenon indirectly indicates the consistency between the criteria we have defined and the reality.}
It can also be clearly observed from the table that \textbf{shallow machine learning-based methods lag behind deep learning-based methods in both classification tasks}.
Although using CodeBERT for embedding combined with machine learning algorithms has achieved good performance, fine-tuning CodeBERT can achieve better performance.
Among the two deep learning models we constructed, \textbf{CodeBERT*2+Bi-LSTM outperformed CodeBERT*2+TextCNN} in overall performance for both classification tasks.
In terms of recall metrics, CodeBERT*2+TextCNN slightly outperforms CodeBERT*2+Bi-LSTM.
This indicates that this method is better at identifying positive samples than distinguishing negative samples.
In the task of classifying TODO forms,  \textbf{CodeBERT*2+Bi-LSTM achieved scores of 94.55\%, 94.18\%, 96.91\%, and 95.51\%} in terms of Accuracy, Precision, Recall, and F1, respectively.
For the task of identifying high-quality TODOs, \textbf{CodeBERT*2+Bi-LSTM can achieve scores of 85.89\%, 85.99\%, 87.71\%, and 86.78\%} in terms of Accuracy, Precision, Recall, and F1, respectively.
This shows that the CodeBERT*2+Bi-LSTM classifiers have achieved a high performance on this dataset.

These experimental results confirm the possibility of using deep learning models to classify TODOs according to form and quality.
We developed a tool for identifying high-quality TODOs based on the CodeBERT*2+Bi-LSTM model at the current stage.
With our tools, developers can discover and improve low-quality TODOs in a timely manner, enabling them to quickly and accurately understand the context when revisiting code in the future.
Moreover, our research on the characteristics of high-quality TODO (Section~\ref{subsec:chara-summ} and Section~\ref{subsec:result-class-todo}) can be utilized to intelligently provide high-quality templates or guidance for developers based on the types of tasks they need to annotate in the future. 

\begin{table}[htbp]
\centering
\caption{The Performance of Different Models in TODO Classification Tasks}
\label{tab:rq3}
\resizebox{\columnwidth}{!}{
\begin{tabular}{lcccccccc}
\hline
\multirow{2}{*}{Models} & \multicolumn{4}{c}{The form of TODOs (Task or Notice)}                    & \multicolumn{4}{c}{The quality of TODOs (Good or Bad)}                    \\ \cline{2-9} 
                        & Accuarcy         & Precision        & Recall           & F1 score         & Accuarcy         & Precision        & Recall           & F1 score         \\ \hline
Part-of-Speech      &    70.38\%  &  67.11\%  &   87.58\%                                &  75.92\%        & 67.80\%  &  64.50\%  &   87.24\%                                &  74.11\%       \\
Multinomial Naive Bayes & 66.50\%  & 70.22\%   & 76.88\%                           & 70.54\%  & 72.22\%  & 76.22\%   & 69.81\%                           & 72.63\%  \\
Logistic Regression     & 86.31\%  & 87.19\%   & 90.32\%                           & 88.25\%  & 80.51\%  & 80.58\%   & 83.44\%                           & 81.46\%  \\
Multi-Layer Perceptron  & 83.83\%  & 85.03\%   & 88.67\%                           & 86.52\%  & 79.88\%  & 79.95\%   & 83.03\%                           & 81.32\%  \\
K-Nearest Neighbors     & 76.84\%  & 78.18\%   & 84.82\%                           & 80.53\%  & 78.07\%  & 78.39\%   & 81.03\%                           & 79.10\%  \\
Random Forest           & 82.78\%  & 83.74\%   & 88.40\%                           & 85.27\%  & 81.38\%  & 80.25\%   & 86.06\%                           & 82.47\%  \\
Gradient Boosting       & 75.27\%  & 73.38\%   & 91.91\%                           & 79.51\%  & 76.60\%  & 74.36\%   & 85.43\%                           & 78.40\%  \\
CodeBERT*2+TextCNN        & 93.69\%       & 92.76\%        & \textbf{97.04\%} &    94.85\%     &   85.48\%       &  82.45\%      & \textbf{88.92\%} &  85.56\%      \\
CodeBERT*2+Bi-LSTM        & \textbf{94.55\%} & \textbf{94.18\%} & 96.91\%          & \textbf{95.51\%} & \textbf{85.89\%} & \textbf{85.99\%} & 87.71\%          & \textbf{86.78\%} \\ \hline
\end{tabular}
}
\end{table}

\find{
{\bf Summary for RQ4:}
We proposed two deep learning-based classifiers to automatically identify a TODO's form (Task or Notice) and quality (Good or Bad). 
Experimental results show the effectiveness of both classifiers on our dataset. 
They can thus be used to proactively mitigate the introduction low-quality TODOs. 
}

\section{Implications}\label{sec:impli}

TODO comments are an important method of managing and coordinating pending coding tasks. 
The quality of TODOs not only affects code implementation, but also has a certain correlation with developers' behavior. 
We discuss key implications for practice and research of our findings in this section.

\subsection{Implications for Practice}

\textbf{Reducing prevalence of low-quality TODOs: } Our study found that 46.7\% of TODOs in open-source software repositories are of low-quality.
A large portion of these low-quality TODOs remain in source files for a long time until resolved or never be formally migrated to change requests.
This finding suggests that TODOs have not fulfilled their intended roles in practice.
For TODOs created by developers, there is a significant need to improve the quality of TODOs and clarify the content of their target tasks. 
This will potentially boost the motivation of developers to resolve TODOs and/or shorten their life cycle.

\noindent\textbf{Developers need better ways for identifying low-quality TODOs: } Our proposed deep learning-based classifiers can help developers identify unqualified TODOs, enabling them to describe specific tasks more clearly or provide more detailed descriptions for developers to understand the potential tasks they need to do next. 
Our tool is able to aid software development practices to include a mechanism, which can remind developers and prompt them to revisit TODOs when low-quality TODOs are identified.

\noindent\textbf{Using high-quality TODOs:} Higher-quality TODOs will help developers more quickly regain context and accurately complete corresponding tasks. They also further
prevent the situation that some TODOs are left in the source code, and eventually become a clean-up or technical debt after several months due to low quality.
Maintaining and increasing  quality over time could assist developers to improve the quality of in-code software documentation. 
We have established connections between TODOs and various software development activities in Section~\ref{subsec:result-class-todo}, while providing a series of key points for creating TODOs. 
These guidelines can serve to significantly improve the quality of TODOs.

\subsection{Implications for Research}

\textbf{Validating TODO study data quality: }There are currently a lot of works~\cite{nie2018natural, nie2019framework, sridhara2016automatically, gao2021automating, tododie, imdone} analyzing and using information from TODOs for academic research and tool development.
However, the data used in these works has not been rigorously screened, leading to the introduction of threats.
For example, if the training set contains many low-quality TODOs with similar content, the actual pending tasks associated with these TODOs may vary widely. Utilizing such a dataset to train a model could cause bias in the output results of the model when inferring. 
This is because, throughout the training process, the model extensively learns the intrinsic correlations between different types of code implementations and similar low-quality TODOs. 
It is very time-consuming to manually filter out these low-quality TODOs.
Our proposed method can achieve an F1 score of 86.87\% in identifying high-quality TODO tasks, which could help researchers clean data and build high-quality datasets.
\changed{
However, in some downstream tasks, it may be necessary to ensure the existence of low-quality TODOs to reflect the reality. 
For tasks where understanding the full scope of TODOs is crucial, researchers can opt to retain these low-quality TODOs to ensure their analysis reflects the authentic state of software development. 
On the other hand, for tasks focused on improving code maintenance or identifying high-impact technical debts, filtering out low-quality TODOs could enhance reliability.
However, viewing it from a positive angle, if our classifier indeed contributes to a decrease in the occurrence of low-quality TODOs within software repositories, thereby improving their overall quality.
This improvement itself represents a form of reality.
}

\noindent \changed{\textbf{Involving low-quality TODOs: }
The tool we proposed performs well in identifying high-quality TODOs.
Consequently, research focusing on low-quality TODOs can also benefit from our tool. 
By efficiently filtering out high-quality TODOs, our tool substantially reduces the effort required to screen for low-quality entries.
Moreover, our in-depth analysis on the characteristics of low-quality TODOs provides valuable insights, which could help the development of tools specifically designed to detect such instances.
As a type of SATD, the characteristics of TODOs may also offer valuable perspectives for addressing other forms of SATD.
For example, by leveraging predefined SATD task categories, it is possible to train models to classify TODOs into different subtypes, thereby enhancing the detection of low-quality comments.
This could help improve the management of some high-priority SATDs.
Meanwhile, low-quality cases within these high-priority SATDs are also more likely to be noticed.
So as to reduce the debt accumulation caused by the long-term existence of these low-quality cases.
}

\noindent\textbf{Further investigating the TODO lifecycle: }
We suggest further exploring the life cycle of TODOs, which includes the timing of TODOs being revisited, kinds of TODOs, and the code suboptimization and technical debt issues potentially caused by the presence of abundance legacy TODOs.

\noindent\textbf{Further improving classifiers and recommenders: }
\changed{Although our tool has shown promising performance in terms of accuracy, precision, recall, and F1 score in our experiments.
However, the direction of improvement for our tools may differ according to specific downstream tasks.
For example, in tasks related to code maintenance and refactoring, high-quality TODOs are important for guiding the direction and priorities of refactoring efforts. 
In this case, accuracy will be even more important. 
Because we aim to minimize errors in labeling low-quality TODO as high-quality, in order to prevent developers from wasting time on unnecessary tasks. 
In project management contexts, the focus may shift towards maximizing recall to ensure that all significant TODOs are identified. In this case, the occurrence of false positives may also be acceptable to a certain extent.}

\changed{Another exciting direction for future work is to build intelligent tools that could automatically suggest high-quality TODO templates or generate recommendations directly within developers' Integrated Development Environments (IDEs).
This would reduce the workload of developers while improving quality. 
In the same time, we will be able to attract more developers to evaluate the tool and utilize their feedback to further improve the performance of our tool.
In this scenario, comprehensive metrics such as accuracy and F1 score become increasingly important. They play a crucial role in boosting developers' trust and ensuring the tool's usability.
}

\section{Threats to Validity}\label{sec:threats}

In this work, we follow the guidelines presented in \cite{wohlin2012experimentation} to address potential threats to the validity of our study. 

\changed{\noindent\textbf{Internal validity} refers to the potential errors in the code implementation. We have rigorously reviewed the source code for our classifiers as well as for the baselines. The Part-of-Speech baseline is easy to implement. Regarding the baselines that employ machine learning algorithms, we utilized the widely recognized open-source software library, scikit-learn~\cite{sklearn}. To optimize the performance of different methods, the parameters in our experiments were carefully fine-tuned. The parameter settings were thoroughly considered and evaluated. Therefore, the impact of this internal threat is very limited.
}

\changed{The second threat to internal validity concerns the potential biases in the TODO lifecycle analysis. According to the matching rules and data processing process mentioned in Section~\ref{subsec:track-method}, there are the following factors that may lead to bias. Firstly, file renaming can lead to missed matches of eliminated TODOs that were previously introduced, thereby reducing the overall count of removed TODOs.
Secondly, the presence of multiple identical TODOs within a single commit may induce mismatches in our matching algorithm.
While manual verification has confirmed these mismatches do not affect our results. But this limitation still needs further improvement. Lastly, we conducted manual inspection to classify removed TODOs as either resolved or unresolved. This process may introduce bias due to differences in inspector's experience. However, collaboration between two examiners has been employed to mitigate this bias as much as possible. In the future, we plan to develop a better matching algorithm (e.g., leveraging file history from version control systems) to enhance the accuracy of TODO lifecycle analysis.}

\changed{Another threat to internal validity relates to the statistical analysis used in our study.
After defining the criteria for high-quality TODOs, we recruited 10 experienced engineers to further validate our labels. We not only calculated the average consistency between the engineers' results and our labels but also computed the Kappa coefficient. This served as a measure of inter-rater agreement among the engineers, thus offering a clear insight into the consistency of our coding process.
Participants received detailed task-specific training and were explicitly forbidden from discussing or assisting each other during the labeling process, which indicates that their results are both credible and independent. Therefore, we consider the threat posed by inter-rater variability to be minimal.}

\changed{In our analysis of the lifecycle of TODOs, significance tests were conducted to evaluate the support for our hypotheses.
We conducted simultaneous tests on four hypotheses using datasets with intersections. As a result, the potential for Type I errors, which occur when a true null hypothesis is incorrectly rejected, could pose a threat to our findings.
Specifically, we applied the Holm-Bonferroni correction to adjust our significance levels.
This correction mitigates the risk of falsely accepting hypotheses due to chance alone.
Through this adjustment, our conclusions are based on statistically sound evidence, thus increasing the reliability of our analysis.
Therefore, the impact of this threat is very limited.
}

\changed{\noindent\textbf{Construct validity} refers to whether the measures utilized in a study can accurately represent the constructs in a real-world context. In our study, the potential threat to this validity is whether the TODOs were extracted and analyzed correctly. To address this issue, we collected 2,863 TODOs from the selected 53 popular GitHub projects based on a set of selective criteria. When constructing our datasets, we did not collect the TODOs from branches except for the master branch. The reason is that other branches may contain a large amount of test code, experimental features, or code that is not yet ready to be merged into the main branch. The process of merging and conflict resolution between different branches may lead to duplicate and incorrect TODO statistics. In addition, code implementations are only merged into the main branch when they are ready and thoroughly tested in many projects. Therefore, the TODOs contained in the master branch better reflect their true lifecycle. Our study either did not consider many other potential data sources, for example small-sized GitHub projects or other collaborative platforms, such as SourceForge. Despite this limitation, we anticipate that the ability of our approach to identify and classify TODOs will enable our results to generalize to other data sources as well. In future work, we will consider studying the characteristics and lifecycle of TODOs in other branches and look to expand our approach to additional data sources to provide a more comprehensive perspective.}

\changed{
Another threat to construct validity relates to the machine-generated TODOs in our dataset.
While preparing our dataset, we did not specifically remove any machine-generated TODOs.
These TODOs only indicate they are auto-generated without providing other useful information, thus they are naturally considered as low-quality TODOs. 
These TODOs either do not affect our investigation into the criteria about TODO quality and the prevalence of different quality TODOs across different projects. 
Furthermore, the characteristics and lifecycle of machine-generated TODOs can also reflect the real-world development practices.
Therefore, the machine-generated TODOs cannot be ignored in RQ1, RQ2, and RQ3.
However, for RQ4, machine-generated TODOs could indeed introduce some bias. 
Due to their repetitive occurrence and distinct features, the classifiers might easily identify them, potentially leading to an inflated performance. 
We identified these TODOs through the keyword “Auto-generated” and quantified their presence. 
In our TODO-introduced dataset which is used to train different classifiers, there were only 49 instances identified as machine-generated TODOs. 
These TODOs only account for 1.7\% of the total dataset.
Therefore, the impact of machine-generated TODOs on our study is considerably minimal.
}

\changed{To ensure high-quality labeling for the automatic classification of TODOs, we established a set of initial criteria and conducted a manual check to determine the labeling criteria more specifically. 
We first randomly sampled 200 pilot TODOs and used the initial criteria to label these samples.
Through this process, the content of the criteria we proposed and their boundaries could be more clearly defined.
It should be noted that the sampling strategy resulted in a 95\% confidence level and a margin of error of 6.68\%. 
Although prior research~\cite{zhou2021finding} has reported findings with a confidence interval similar to ours, it's noteworthy that their selection was based on a 95\% confidence level with a margin of error of 10.0\%. This factor may still pose a potential threat to the reliability of our criteria.
Nevertheless, our final proposed criteria was derived from an extensive survey of gray literature, thorough paper reviews, and careful examination of the samples.
During the labeling process, all the authors discussed the discrepancies and reached an agreement on data labeling. 
Furthermore, to enhance the reliability of our criteria, we recruited 10 experienced developers to validate our labels, ensuring our criteria are both robust and grounded in practice.
Therefore, the threat has been mitigated as much as possible.
For the formal data labeling, we conducted two rounds of manual data labeling between the first two authors and discussed the conflicts with the third author. The validation from developers ensures the quality of data labeling. Finally, we used all the 2,863 labeled TODO-introduced data to train the classification models. However, we acknowledge that there might still be some bias on data labeling. 
In the future, we plan to incorporate additional data sources and evaluators to further validate the reliability of our criteria and the accuracy of our labels.}

\changed{We described the thematic analysis process in Section~\ref{subsec:chara-summ} to investigate the characteristics of TODOs. 
Should the generated codes fail to accurately represent the targeted concepts, or if the volume of sampled data is insufficient for a thorough exploration, either situation could potentially introduce threats to the validity of our study.
Our coders have carefully reviewed the collected TODOs and TODO-related literature, and they possess substantial knowledge in the field of software engineering. These all help ensure that the codes generated during the thematic analysis process accurately reflect the expected research concepts.
To mitigate the second threat, we sampled a total of 600 TODOs for investigation, with a margin of error of 4.68\% at a 95\% confidence level. Therefore, the potential threats to construct validity posed by this process is limited.}

When estimating the classifiers, we used the processed TODO-introduced dataset with 2,863 TODO data pairs.
The size of this dataset is not large enough for deep learning-based classifiers.
It may introduce bias and affect the representativeness of the experimental results.
Although we applied 10-fold cross-validation to minimize the impact of this threat, expanding the coverage of the dataset is still one of the main focuses of our future work.

\noindent\textbf{External validity} concerns the generalisation of the study particularly regarding the comprehensiveness of our TODO data. To address this threat, we collected the data by examining most popular Java GitHub projects, which are in different domains. To automate the classification tasks, we utilized the pre-trained CodeBERT model, which has demonstrated optimal performance in recent studies \cite{gao2020technical}. However, it should be noted that our results only apply to the specific programming language projects (i.e., Java) and deep learning model used in this study and do not provide insight into the effectiveness of employing other language projects and models with different structures or advanced features.

\noindent\textbf{Reliability} pertains to the consistency of study results when replicated by other researchers. In this study, potential threats to reliability stem from the collection and analysis processes of the TODO data. However, we have addressed these concerns by providing a clear and explicit description of our data collection and analysis procedures. 

\section{Conclusion}\label{sec:conclusion}
In this paper, we presented a multi-method study that investigates the quality of TODOs in open-source software projects and its impact.
Our analysis of a large number of TODOs reveals a considerable proportion of low-quality TODOs in software repositories, which cause extra efforts and time for software development and maintenance.
We summarized a set of criteria for identifying high-quality TODOs and described how to use TODOs to support a variety of articulation activities within software development.
Based on our findings, we constructed deep learning models for classifying TODOs, while serving as the cornerstone for future intelligent tool development.
At last, we investigated the life cycle of TODOs and found that different quality TODOs have diverse management and maintenance characteristics.
These findings further demonstrate the significance of our research and the potential value of proposed classifiers.
Our research findings can provide guidance for developers when writing TODOs and inspire researchers in the field of software documentation.

\section*{Acknowledgements}

This research was partially supported by the National Natural Science Foundation of China (No. 62302430 and No. 62202341), Zhejiang Provincial Natural Science Foundation of China (No. LQ24F020017), ARC Laureate Fellowship (FL190100035), Zhejiang Province “JianBingLingYan+X” Research and Development Plan (2024C01114), the Joint Funds of the Zhejiang Provincial Natural Science Foundation of China under Grant No. LHZSD24F020001. 
This work was also supported by the Starry Night Science Fund of Zhejiang University Shanghai Institute for Advanced Study, Grant No. SN-ZJU-SIAS-001 and Shanghai Sailing Program (23YF1446900). 
This work is supported by Zhejiang Provincial Engineering Research Center for Real-Time SmartTech in Urban Security Governance. 
The numerical calculations in this paper have been done on the supercomputing system in the Supercomputing Center of Hangzhou City University.

\bibliographystyle{ACM-Reference-Format}
\bibliography{sample-base}

\end{document}